\documentclass[%
 aip,
 amsmath,amssymb,
 reprint,%
]{revtex4-1}

\usepackage{graphicx}
\usepackage{dcolumn}
\usepackage{bm}

\usepackage[utf8]{inputenc}
\usepackage[T1]{fontenc}
\usepackage{mathptmx}
\usepackage{etoolbox}

\usepackage{hyperref}
\usepackage{cleveref}
\usepackage{graphicx}
\usepackage{dcolumn}
\usepackage{bm}
\usepackage[caption=false]{subfig}
\usepackage{float}
\usepackage{xcolor}
\usepackage{lipsum}
\newcommand{\R}{\mathbf{R}}

\usepackage{mathtools, nccmath}

\makeatletter
\def\@email#1#2{%
 \endgroup
 \patchcmd{\titleblock@produce}
  {\frontmatter@RRAPformat}
  {\frontmatter@RRAPformat{\produce@RRAP{*#1\href{mailto:#2}{#2}}}\frontmatter@RRAPformat}
  {}{}
}%
\makeatother

\begin{document}

\preprint{AIP/123-QED}


\title{Active dynamics of charged macromolecules}
\author{Tapas Singha} 
\email{tapas134@gmail.com}
\affiliation{Laboratoire Physique des Cellules et Cancer (PCC), Institut Curie, PSL Research University, CNRS UMR168, 75005 Paris, France}
\author{Siao-Fong Li}
\affiliation{Department of Physics, University of Massachusetts, Amherst, Massachusetts 01003, USA}
\author{Murugappan Muthukumar}
\affiliation{Department of Polymer Science and Engineering, University of Massachusetts, Amherst, Massachusetts 01003, USA}
\date{\today}

\begin{abstract}

We study the role of active coupling on the transport properties of homogeneously charged macromolecules in an infinitely dilute solution. An enzyme becomes actively bound to a segment of the macromolecule, exerting an electrostatic force on it. Eventually, thermal fluctuations cause it to become unbound, introducing active coupling into the system. We study the mean-squared displacement (MSD) and find a new scaling regime compared to the thermal counterpart in the presence of hydrodynamic and segment-segment electrostatic interactions. Furthermore, the study of segment-segment equal-time correlation reveals the swelling of the macromolecule. Further, we derive the concentration equation of the macromolecule with active binding and study how the cooperative diffusivity of the macromolecules get modified by its environment, including the macromolecules itself. It turns out that these active fluctuations enhance the effective diffusivity of the macromolecules. The derived closed-form expression for diffusion constant is pertinent to the accurate interpretation of light scattering data in multi-component systems with binding-unbinding equilibria.
\end{abstract}
\maketitle

\section{Introduction}

A widespread biological system consists of large molecules, such as nucleotides, proteins, and enzymes etc, which carry electric charges. The dynamics of these charged macromolecules in solution is typically governed by various physical and chemical driving forces, affecting their transport properties. Many biophysical processes, including the dynamics of molecules due to protein binding, the interaction of myosin with actin filaments in a cytoskeleton network\cite{Howard_2001}, remodeling of chromatin by ATPase \cite{Cell_Narlikar_2013}, strongly depend on the transport properties of the macromolecules. In addition, there has been a significant use of microdevices and nanomedicine \cite{ACSNano_Wang_2012, NanoLett_Sen_2015, NanoLett_Medina_2016} for biomedical applications such as drug delivery. Understanding and controlling the dynamic properties in these synthetic polymeric systems are also essential requirements.

In the last two decades, there has been a growing interest in the field of macromolecules that use energy from ATP hydrolysis. These systems involve macromolecules composed of either active building blocks \cite{PRE_Aitor_2020} or passive polymers immersed in a solution containing self-propelled active species \cite{JCP_Eisenstecken_2017, JPSJ_Winkler_2017}. A considerable number of theoretical and numerical studies have been carried out, ranging from studies of the dynamics of individual polymers \cite{JCP_Dino_2018, JCP_Kaiser2014, PRE_Harder_2014} to investigating collections of polymers subjected to active coupling \cite{RMP_Marchetti_2013}. These studies have explored various aspects of polymer behavior, including activity-induced swelling \cite{JCP_Kaiser2014, PRL_Bianco_2018, JCP_Chaki_2019}, aggregation \cite{SoftMatter_Isele_2015}, and pattern formation \cite{Chelakkot_2013}.

A theoretical study of self-propelled colloidal particles demonstrates that diffusion constant increases due to self-electrophoresis \cite{GolestanianPRL2009}, while experiments with catalytic enzymes show an enhancement of diffusivity through electrophoresis \cite{MuddanaJACS2009, Sengupta2013}. In other theoretical studies, a neutral flexible polymer in an environment of active enzymes (energized by ATP) is considered, where enzymes generate dipolar forces along the polymer backbone upon binding \cite{Put_PRE2019}. Experiments on the transcription of RNA polymerase on a fluorescently labeled DNA template indicate a significant enhancement in the diffusion constant of DNA \cite{Yu_JACS2009}. Though the diffusion constant of enzyme has been studied in the presence of substrate, investigations into the effect of enzymes on macromolecules have yet to be explored.

However, based on our current knowledge, polymer studies often overlook explicitly considering charged species and their electrostatic interactions as a source of activity, despite the fact that most cellular components are charged and induce active coupling upon binding without ATP hydrolysis. This form of active coupling differs fundamentally from typical active particles, which usually derive energy from ATP hydrolysis and dissipate energy through mechanical force exertion \cite{RMP_Marchetti_2013}.

On the other hand, a significant area of research in the field of charged macromolecules focuses on studying the dynamic properties of charged polymeric solutions through numerous theoretical, numerical, and experimental works \cite{Muthukumar2016, Parker1997}. Despite the overall electrical neutrality of the solution, the dynamics of a charged polymer within it are strongly coupled by the electrostatic interactions of the dissociated counterions.

In this view, while the ATP hydrolysis-mediated active dynamics of macromolecules have been extensively studied in active matter, electrostatic-driven macromolecular dynamics have also received significant attention. Nevertheless, it has not yet been realized that electrostatic interactions may induce active coupling without ATP hydrolysis in these charged systems, which could bridge the gap between these two growing fields.

To thoroughly understand the impact of active coupling on polymer dynamics, we study the behavior of macromolecules by calculating static segment-to-segment correlations and mean-squared displacement of the segments. Incorporating hydrodynamic and segment-segment interactions, we introduce active coupling and investigate its effects on both static and dynamic correlations. Within segment-segment interactions, we account for excluded volume effects and electrostatic interactions among all possible pairs of segments.

We aim to understand how a charged macromolecule moves in a  diluted solution with active coupling. To do this, we  consider the mobility of the polymer with counterions and the friction between charged segments and solvent \cite{MuthuACP2005}. We follow the Fokker-Planck formalism with colored noise \cite{Hanggi_1995} and employ it to find the concentration equation of a polymer in the presence of enzymes. By studying the time evolution of counterion concentration, we derive the cooperative diffusivity of the polymer in large scale limit which can be measured in the dynamic light scattering experiment. We find that the dynamics of the macromolecule, mediated by counterions, get significantly affected when enzymes bind and unbind, enhancing the cooperative diffusivity of the charged polymer. Our closed-form expression for cooperative diffusivity clearly demonstrates how electrostatic force, temperature, and screening length impact diffusivity. We believe that the systematic understanding of enhancement of the diffusivity of macromolecules may have broad applicability in terms of controlling the transport properties in the biophysical systems. 

We organize our work as follows. In Section II, we define the model describing the basic features of the system. In Section III, we present the time-independent segment-segment correlation and mean-squared displacement for the model. Moving on to Section IV, we focus on deriving the cooperative diffusivity of a macromolecule. Finally, in Section VII, we summarize our results and conclude our work.

\section{Model}
We study the dynamics of a homogeneously charged polymer in an infinitely dilute solution where hydrodynamic interactions become important. In particular, we aim to understand dynamics of a charged polymer in a dilute solutions. The Langevin equation for a charged polymer segment of position vector $\mathbf{R}(s,t)$ driven by thermal noise $\mathbf{f}_T $ and the force of actively binding and unbinding of enzymes $\mathbf{f}_{A}$, can be expressed as
\begin{eqnarray}
&&\frac{\partial \mathbf{R}(s,t)}{\partial t} = \int^{N}_{0} ds' \, \mathbf{H} (\mathbf{R}(s)-\mathbf{R}(s')) [ \frac{3 k_B T}{\ell^2} \frac{\partial^2 \mathbf{R}(s',t)}{\partial s'^2} \nonumber \\ && + \frac{\partial }{\partial \mathbf{R}(s',t)} V[\mathbf{R}(s)-\mathbf{R}(s')] + \mathbf{f}_T (s', t)  + \mathbf{f}_{A}(s', t) ]
\label{GAZ}
\end{eqnarray}
where $s$ denotes the index of segment along the contour of the polymer, and $\ell$ represents the Kuhn length. The first term on the right-hand side of Eq.\,(\ref{GAZ}) accounts for the connectivity of the $s'$th segment where $T$ and $k_B$ denote the temperature and Boltzmann constant, respectively. Oseen tensor $\mathbf{H}$ accounts for the intra-chain hydrodynamic interactions which can be expressed as
\begin{align}
\label{OseenTensor}
\mathbf{H} \left(\left|\mathbf{R}_i(s)-\mathbf{R}_i(s^\prime)\right|\right)=
\begin{cases}
\frac{\delta(s-s^{\prime})}{\zeta},&\text{ $s=s^\prime$ }\\
&\\
\frac{1+\hat{\mathbf{r}}_{s,s^\prime}\hat{\mathbf{r}}_{s,s^\prime}}{\frac{4\zeta}{3}\left|\mathbf{R}_i(s)-\mathbf{R}_i(s^\prime)\right|}, &\text{ $s\neq  s^\prime$ }
\end{cases}.
\end{align}
where $\hat{\mathbf{r}}_{s,s^\prime}$ denotes the unit vector $\mathbf{R}_i(s)-\mathbf{R}_i(s^\prime)/\left|\mathbf{R}_i(s)-\mathbf{R}_i(s^\prime)\right|$.
The second term represents the electrostatic and excluded volume interaction between the $s'$th segment and other segments of the polymer, both of which are highly nonlinear. The form of potential can be expressed as 
\begin{equation}
\frac{V}{k_B T} = w\,\, \delta (\R(s)-\R(s')) + \frac{z^2_p \ell_{B}}{|\R(s) -\R(s')|} \,\, e^{-\kappa \, |\R(s) -\R(s')|}
\label{ExcludedPlusElectro}
\end{equation}
where $w$ is the coefficient of the excluded volume interaction, and $z_p$ is the valency of a charged segment. In the above equation, $\ell_{B}=e^2/(\epsilon k_B T)$ is the Bjerrum length, where $\epsilon$ is the effective dielectric constant of the solution. The parameter $\kappa$ is the inverse Debye length, which determines the electrostatic screening length due to dissociated counterions and salt ions in the solvent.

\begin{figure}[t]
\centering
\begin{minipage}{0.46\textwidth}
    \includegraphics[width=\textwidth,height=0.27\textheight]{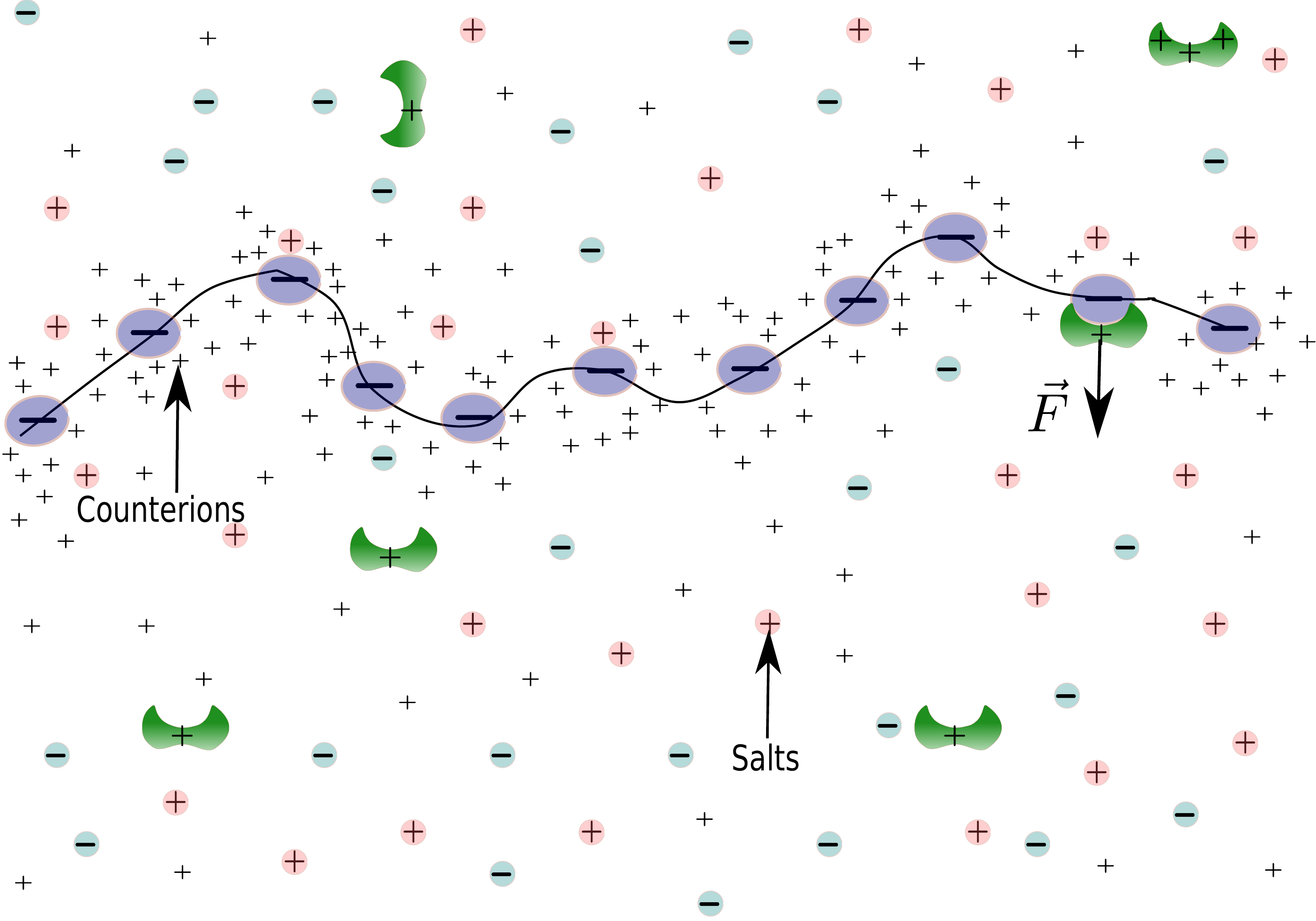}
    \end{minipage}%
 \caption{\label{ChargedPolymerEnzyme} A schematic diagram of a polyelectrolyte solution with the charged enzyme and salt ions. The negatively charged macromolecule is surrounded by its positively charged dissociated counterions.}
 \label{schematic}
\end{figure}

\subsection{Autocorrelation of Telegraphic force}
In reality, enzymes may bind multiple monomers of a biopolymer. For simplicity, we coarse-grain the multiple monomers into a segment, and we consider an enzyme binds to a segment. The term $\mathbf{f}_A$ in Eq.\,(\ref{GAZ}) represents the force acting on a segment because of an enzyme binding and unbinding to a segment (shown in Fig.\,\ref{schematic}).  An enzyme switches between being bound and unbound, or vice versa. This transition corresponds to binding force $\mathbf{f}_{+}$ and unbinding force $\mathbf{f}_{-}=\mathbf{0}$.
When $A'$ is in the initial state at time $t_0$ (where $A'$ can be either bound ($+$) or unbound $(-)$), it evolves into new states: bound, with transition probability $P(+,t|A', t_0)$, and unbound, with $P(-,t|A', t_0)$. The Master equation reads as
\begin{equation}
\frac{\partial P(i,t|A', t_0)}{\partial t} = -\lambda_{i} \, P(i,t|A', t_0) + \lambda_{j} \, P(j,t|A', t_0)
\end{equation}
where $i \neq j$, and $i$ and $j$ have two possible states as $+/-$. The steady-state probabilities of bound and unbound states can be obtained as $P_{b} = \tau_{b}/(\tau_{b}+\tau_{ub}), \,\, P_{ub} = \tau_{ub}/(\tau_{b}+\tau_{ub}) $
where $\lambda_{-}=1/\tau_{ub}$ and $\lambda_{+}=1/\tau_{b}$. When a single enzyme and a segment of a polymer participate in the binding and unbinding process, each bound event is followed by an unbound event, and in an unbound state, the amplitude is zero (i.e., $\mathbf{f}_-=0$). Considering
the kinetics of the bound and unbound dynamics of the enzyme as a telegraphic process \cite{gardiner2004handbook}, we derive the autocorrelation of telegraphic force (the details are given in the \textbf{Appendix}) as
\begin{equation}
  \begin{split}
\langle \mathbf{f}_{A} (s, t) \cdot \mathbf{f}_{A}(s', t')\rangle  =  \,P^2_{b} \, |f|^2 \left[ 1  + p\, e^{-|t-t'|/\tau'}\right] \\ \times e^{-|t-t'|/\tau_{\theta}}   \delta (s-s')
\label{AutoActiveII}
\end{split}
\end{equation}
where we consider $f^2=f_{+}^2$ and $\theta$ is the angle between the force $f_A(t)$ at time $t$ and  $f_A(t')$ at other time $t'$, where $1/\tau'= 1/\tau_{ub}+1/\tau_{b}$, $p=\tau_{ub}/\tau_{b}$, and $\tau_{\theta}$ is the timescale of rotational diffusion of the enzyme. As we can see the orientations of the binding forces introduce another layer of randomness because of thermal fluctuations.

\subsection{Fourier transform}

We consider Fourier transform of the position vector of a charged segment as 
$ \R(s,t) = \int^{\infty}_{-\infty} \frac{dq}{[2\pi]} \int^{\infty}_{-\infty} \frac{d \omega}{[2\pi]} \,\,  e^{i(q s+\omega t)}  \widehat{\R}(q,\omega)$ and $\quad \widehat{\R}(q,t) = \int^{N}_{0} ds \, e^{-i q\,s} \, \R(s,t) $ where $q$ is the mode number and $\omega$ is the frequency in Fourier space. Let us separated out the center of mass (cm) part ($q=0$) from the rest of the modes. We first take the Fourier transform of Eq.\,(\ref{GAZ}), and consider pre-averaging  and uniform expansion approximations \cite{JCP_Muthu1982} for $H(|\R(s)-\R(s')|)$ which yields
\begin{equation}
 \frac{\partial \widehat{\R}(q, t)}{\partial t} = -\,H(q)[\frac{3k_B T}{\ell\, \ell_{\text{eff}}(q)} q^2 \widehat{\R}(q,t) + \widehat{\mathbf{f}}_{T} (q,t) + \widehat{\mathbf{f}}_A (q,t)]
 \label{EqGenActiveZimm}
\end{equation}
for $q\neq 0$. We combine several parameters of the polymer as follows: $ \tau_{1} =\zeta \ell^2/3 k_{B}T$ where $\zeta$ is the frictional coefficient.  The segment-segment electrostatic interactions \cite{Muthukumar1987} are captured by $\ell_{\text{eff}}(q)$, and we consider $ \ell_{\text{eff}}(q)/\ell = 1 + a_{\ell} \, \, |q|^{1-2\nu}
\simeq \lim_{q\to 0} \,\, a_{\ell} \, \, |q|^{1-2\nu}$ similarly, $\zeta H(q) = 1 + a_h \,\, |q|^{\nu-1} \simeq \lim_{q\to 0} a_h \,\, |q|^{\nu-1}$
where $\nu$ is the size exponent of the macromolecule, and $a_{\ell}$ and $a_{h}$ are the dimensionless parameters related to electrostatic and hydrodynamic interaction, respectively. In Eq.\,\ref{EqGenActiveZimm}, the relaxation timescale of a single segment can be written as $ \tilde{\tau}_1(q)=(\ell\, \ell_{\text{eff}})/(3k_B T H(q))= \overline{\tau}_1 |q|^{2-3\nu}$
where $\overline{\tau}_1 = \tau_{1}a_{\ell}/a_{h}$. Using the Fourier transform, we calculate the auto-correlation of thermal noise as $\langle \mathbf{\hat{f}}_{T}(q,\omega) \cdot \mathbf{\hat{f}}_{T}(q', \omega')\rangle = \left(6 k_B T/H(q)\right)\,(2\pi)^2 \delta(q+q')  \delta(\omega+\omega')$
and similarly the auto-correlation of the active-force is obtained as
\begin{eqnarray}
\langle \hat{\mathbf{f}}_A(q, \omega) \cdot \hat{\mathbf{f}}_A(q', \omega')  \rangle =8(\pi P_{b}&&|f|)^2 \biggl(\frac{\tau_{\theta}}{1+(\omega\tau_{\theta})^2}+  \frac{p\,\tau}{1+ (\omega\,\tau)^2} \biggr) \nonumber \\ && \delta(q+q') \delta(\omega+\omega')
\label{EqFT_AutoCorr}
\end{eqnarray}
where $1/\tau = \tau_b^{-1} + \tau_{ub}^{-1} + \tau_{\theta}^{-1}$.


\subsection{Segment-Segment Correlation function}
We study the segment-segment correlation and mean-squared displacement (MSD) of a segment with the active coupling. We first write the segment-segment correlation function in Fourier-space as 
\begin{eqnarray}
&& \resizebox{0.14\textwidth}{!}{$\langle [\mathbf{R}(s,t)-\mathbf{R}(s',t')]^2  \rangle$} = \medint\int\medint\int\medint \int\medint\int\frac{dq\,dq'd\omega\,d \omega'}{(2\pi)^4}  \resizebox{0.14\textwidth}{!}{$\langle \mathbf{R}(q,\omega) \cdot \mathbf{R}(q',\omega')  \rangle$} \nonumber \\
&& \times [ e^{i(q+q')s+i(\omega+\omega')t} + e^{i(q+q')s'+i(\omega + \omega')t'}  
- 2\, e^{i(q s + q' s')+i(\omega t+ \omega' t')}] \nonumber \\
\label{EqStrucFuncMain}
\end{eqnarray}

Expressing the correlation in $q$ and $\omega$ space, we first calculate  $\resizebox{0.14\textwidth}{!}{$\langle \R (q,\omega) \cdot \R(q',\omega') \rangle$} $, and perform the integration over $\omega'$ and $q'$, we obtain 
\begin{eqnarray}
\resizebox{0.15\textwidth}{!}{$\langle[\mathbf{R}(s,t)-\mathbf{R}(s',t')]^2\rangle$} && = 2 \medint\int^{\infty}_{-\infty} \frac{dq}{(2\pi)}\medint\int^{\infty}_{-\infty}\frac{d\omega}{(2\pi)} 
\bigl( 6 k_B T H(q)  \nonumber \\ &&  +2 |f|^2\, P^2_b \, H^{2}(q)  (\frac{\tau_{\theta}}{1+(\omega\tau_{\theta})^2}+ \frac{p\, \tau}{1+(\omega\tau)^2}\bigr) \nonumber\\ && 
\times \frac{[1-\cos(|s-s'| q)e^{-i\omega (t-t')}]}{\omega^2+ (\frac{q^2}{\tilde{\tau}_1(q)})^2} - \Delta (q_{\text{low}})
\label{MonomerDisplaceWholeMain}
\end{eqnarray}
where term $\Delta (q_{\text{low}})$ is the contribution of the lower modes in the limit $q \rightarrow 0$. It is important to note that here we  focus on studying the effect of all modes except $q=0$ which corresponds to the center of mass. Interestingly, the segment-segment correlation can be estimated experimentally by measuring the dynamical structure factor. Using the definition of the dynamical structure factor as $
 S(\mathbf{k}, t) = \exp{\left(-\frac{k^2}{6} \langle [\mathbf{R}(s,t) - \mathbf{R}(s,t')]^2  \rangle_{\text{total}} \right) } .$ One can calculate $S(\mathbf{k}, t)\sim \exp{(-k^2 |t-t'|^{f(\nu)})}$ where $k$ is the scattering wave vector which
can be measured in incoherent inelastic neutron scattering via decay rate $r_{d}\sim k^{2/f(\nu)} $. Further, we present the results of equal-time segment-segment correlation and mean-squared displacement.
\section{Results}
\subsection{Equal time segment-segment correlation}
Here we focus on the static properties of the polymer in the presence of active coupling, hydrodynamics and electrostatic interactions. We aim to calculate the 
 equal-time segment-segment correlation, and thereby end-to-end distance. (More detail of the derivation of the correlation is given in the Appendix.) \paragraph{Thermal contribution:}  Substituting $t=t'$ in Eq.\,(\ref{MonomerDisplaceWholeMain}), we first calculate the thermal contribution as  
\begin{eqnarray}
\langle[\R(s,t)-\R(s',t)]^2\rangle_{\text{T}}
\simeq B_{T} \,\, |s-s'|^{2\nu} 
\label{FTGenCorrThermalMain}
\end{eqnarray}
where the subscript $T$ stands for thermal fluctuations. The information of the electrostatic and hydrodynamic interactions are buried in $a_{\ell}$, and the coefficient $B_T = (2\,a_{\ell}\,\ell^2/\pi)\,  \cos(\pi (1+\nu)) \, \Gamma(-2\nu).$
\paragraph{Active contribution:} Substituting $t=t'$, we perform the integration given in Eq.\,(\ref{MonomerDisplaceWholeMain}) for the active part, and obtain the correlation between two segments of a single polymer resulting from the active coupling (further details can be found in the appendices).  
First, we derive the active correlation when the separation of the indices of the two segments much less than the ratio of the relaxation time $\tau$ and single segment relaxation time $\tau_1$. In the limit $|s-s'|\ll (\tau/\overline{\tau}_1)^{1/3\nu}$, the correlation is obtained as 
\begin{eqnarray}
\langle [\R(s,t)-\R(s',t)]^2 \rangle_{\text{A}} \simeq 
B^{1}_{A} \,\,  |s-s'|^2  
\end{eqnarray} 
where 
\begin{eqnarray}
B^{1}_{A} = \frac{2 b(\nu)}{\pi} \left(P_b |f|\, \frac{a_h \overline{\tau}_1}{\zeta}\right)^2\,  \left(\frac{\tau_{\theta}}{\overline{\tau}_1}\right)^{\frac{4}{3}-\frac{1}{3\nu}} [1+p \left(\frac{\tau}{\tau_{\theta}}\right)^{\frac{4}{3}-\frac{1}{3\nu}}]\nonumber\\
\end{eqnarray} 
and $b(\nu)=(1/6\nu) \Gamma(1/3\nu-1/3) \Gamma(4/3-1/3\nu)$. When the relative index between two segments is smaller, the contribution from the active part varies rod-like $|s-s'|^2$. This indicates the swelling of the polymer in  smaller scales.  Recalling $1/\tau = \tau^{-1}_{b}+\tau^{-1}_{ub}+\tau^{-1}_{\theta}$, therefore, $\tau_{\theta}> \tau$, for $\quad |s-s'| \gg \bigl(\tau_{\theta}/\overline{\tau}_1\bigr)^{1/3\nu}$, we obtain the correlation as 
\begin{eqnarray}
\langle [\R(s,t)-\R(s',t)]^2 \rangle_{\text{A}} \simeq
B^{2}_{A} \,\, |s-s'|^{1+\nu}   
\end{eqnarray}
where 
\begin{eqnarray}
B^{2}_{A} = \frac{2g(\nu)}{\pi} \, \left(\frac{P_b |f|\, a_h \overline{\tau}_1}{\zeta}\right)^2  \,\left(\frac{\tau_{\theta}}{\overline{\tau}_1}\right) \, \left(1+\frac{p\tau}{\tau_{\theta}}\right)
\end{eqnarray}
and $g(\nu)=\sin(\pi \nu/2)/(\pi (1+\nu))\,  \Gamma(\frac{1-\nu}{2}).$ There is a crossover regime where  $(\tau/\overline{\tau}_1)^{1/3\nu} \ll |s-s'| \ll (\tau_{\theta}/\overline{\tau}_1)^{1/3\nu}$ where both the exponents of $|s-s'|$ exists. Note that we derive general expressions for correlation in terms of size exponent $\nu$ for the correlation, and active binding induces a persistence lengths $(\tau/\overline{\tau}_1)^{1/3\nu}$ and $\bigl(\tau_{\theta}/\overline{\tau}_1\bigr)^{1/3\nu}$ for the above two cases, respectively. The polyelectrolyte is rod-like at a short scale, at a large scale, the binding force also expands the polyelectrolyte with a scaling behavior $|s-s^\prime|^{1+\nu}$ which is weaker than that at a short scale. Once the expanding effect of the binding force dominates, it affects the pre-averaging approximation as the hydrodynamic interactions depend on conformations. Therefore, we consider the weak binding force $f$ which may also arise from a long unbound time $\tau_{ub}$, or a short rotation time $\tau_\theta$.

The shape of a polymer can be characterized by the average of the mean-squared of the end-to-end distance. We find $\langle [\R(N,t)-\R(0,t)]^2 \rangle_{\text{total}} \simeq B^{2}_{A} \, N^{1+\nu} $ for large $N$. With hydrodynamic interactions, the active force varies as $N^{1+\nu}$, whereas the contribution from the thermal force varies as $N^{2\nu}$. Thus, for large $N$, we find that the active force dominates over the thermal noise. Note that in the asymptotic limits, we obtain the scaling of the active parts with their prefactors, and the results corresponding to the thermal part are obtained exactly. 
 

\subsection{Mean-squared displacement}
Further, we analytically study the mean-squared displacement of a segment with hydrodynamic and segment-segment electrostatic interactions. 
\paragraph{Thermal contribution:}
Substituting $s=s'$ in Eq.\,\ref{MonomerDisplaceWholeMain}, we first determine the contribution to the thermal part for a single segment expressed as:
\begin{align}
\langle [\R(s,t)-\R(s,t')]^2  \rangle_{\text{T}} &=C_T\, |t-t'|^{\frac{2}{3}}
\end{align}
where 
\begin{equation}
C_T= \frac{\Gamma \left(\frac{1}{3}\right)}{\pi \nu}\, \frac{3 k_B T\,a_h}{\zeta}\, \overline{\tau}^{1/3}_1.
\end{equation}

\paragraph{Active force contribution:} Due to the active coupling, various regimes emerge as $|t-t'|$ varies from short to long times, such as small times $|t-t'|\ll \tau$, and secondly, for times $\tau_{\theta}\ll |t-t'|\ll \tau_{\text{Zimm}}$. First, in the regime, $|t-t'|\ll \tau$, MSD reads as
\begin{align}
\langle [\R(s,t)-\R(s,t')]^2 \rangle_{\text{A}}
\simeq\, C^{1}_A \,\,  |t-t'|^2 
\end{align}
when $|t-t'|\ll \tau $ where 
\begin{equation}
C^{1}_A =d(\nu)\,\left(\frac{P_b |f|\, a_h }{\zeta}\right)^2 \,\,\overline{\tau}^{\frac{2}{3}-\frac{1}{3\nu}}_1\, 
  ( \tau_{\theta}^{\frac{1}{3\nu}-\frac{2}{3}} +p\, \tau^{\frac{1}{3\nu}-\frac{2}{3}}).
\end{equation}
and $d(\nu)=\Gamma \left(\frac{2}{3}-\frac{1}{3\nu}\right) \Gamma(\frac{1}{3}+\frac{1}{3\nu})/(3\pi \nu).$ In the short time limit, i.e., when the observation time $|t-t'| \ll \tau$, mean-squared displacement of  a tagged segment shows ballistic behavior as it varies with $|t-t'|^2$. Interestingly, in the limit $|t-t'| \gg \tau_{\theta}$, the contribution of active coupling to MSD also deviates from the scaling of the thermal force which is obtained as
\begin{align}
\langle [\R(s,t)-\R(s,t')]^2 \rangle_{\text{A}}
\simeq  C^{2}_A \,\, |t-t'|^{\frac{1}{3}+\frac{1}{3\nu}}
\end{align}
where 
\begin{equation}
C^{2}_A =\frac{2\Gamma(\frac{2}{3}-\frac{1}{3\nu})}{\pi (1+\nu)}\,
\left(\frac{P_b |f|\, a_h }{\zeta}\right)^2 \,\,\overline{\tau}^{\frac{2}{3}-\frac{1}{3\nu}}_1\,  \left(\tau_{\theta}+p\, \tau \right).
\end{equation}
There is a crossover regime when $ \tau \ll |t-t'| \ll \tau_{\theta}$ where we get the transition from one  to other regime. The obtained scaling form is general for any feasible size exponent $\nu$. In our preliminary attempt, we study the mean-squared-displacement with and without hydrodynamic interactions \cite{Singha2021}.

\subsection{Mean squared displacement of center of mass}
We write the Langevin equation for center of mass as
\begin{equation}
f_r \frac{\partial }{\partial t} \widehat{\mathbf{R}}_0(t) = \widehat{\mathbf{f}}_T (0, t)  + \widehat{\mathbf{f}}_{A}(0, t)  
\label{TimeEvoSeg_FT}
\end{equation}
where $\widehat{\mathbf{R}}_0(t) = \int^{N}_{0} ds\, \mathbf{R}(s,t)$. The effect of solvent and segment-segment interactions is captured by $f_r$. The second and third terms, $\mathbf{\hat{f}}_T (0, t)$ and $\mathbf{\hat{f}}_{A}(0, t)$, represent the total thermal noise and active force experienced by the entire polymer $\langle \widehat{\mathbf{f}}_T (0, t) \widehat{\mathbf{f}}_T (0, t') \rangle = 6k_B T N f_r \delta(t-t')$ while for active correlation is obtained as
\begin{align}
\langle \widehat{\mathbf{f}}_A (0, t) \cdot \widehat{\mathbf{f}}_A (0, t') \rangle = N P^2_{b} |f|^2 \bigl[ 1  + p e^{-|t-t'|/\tau'} \bigr] e^{-|t-t'|/\tau_{\theta}} 
\label{ActiveForceCorrCM}
\end{align}
where $p= P_{ub}/P_{b} = \tau_{ub}/\tau_{b}$. Further, we discuss the equations for the center of mass (cm) of the polymer, defined as $ \mathbf{R}_{\text{cm}}(t) = \frac{1}{N} \int^{N}_{0} \mathbf{R}(s,t) $, where $\widehat{\mathbf{R}}_0(t) = N \mathbf{R}_{\text{cm}}(t)$. The dynamics of the center of mass of the polymer can be calculated as
\begin{equation}
\begin{split}
\resizebox{0.15\textwidth}{!}{$\langle [\mathbf{R}_{\text{cm}}(t)-\mathbf{R}_{\text{cm}}(t_0)]^2 \rangle$} =(\frac{6k_B T}{N f_{r}}+ \frac{2P_b^2|f|^2}{N f_{r}^2}(\tau_{\theta}+p\tau)) |t-t_0|  \\  + 2 \frac{P_b^2|f|^2}{N f_{r}^2} [\tau^2_{\theta} (e^{-|t-t_0|/\tau_{\theta}}-1)+p\tau^2\,( e^{-|t-t_0|/\tau}-1)]
\end{split}
\end{equation}
In the limit $|t-t_0| \ll \tau$, the mean square displacement (MSD) of the center of mass can be expressed as
\begin{equation}
\resizebox{0.15\textwidth}{!}{$\langle [\mathbf{R}_{\text{cm}}(t)-\mathbf{R}_{\text{cm}}(t_0)]^2 \rangle$} = \frac{6k_B T}{N f_{r}} (t-t_0) + (1+p) \frac{P_b^2|f|^2}{N f_{r}^2} (t-t_0)^2
\end{equation}
On the other hand, in the limit $|t-t_0| \gg \tau_{\theta}>\tau$, we get $ \resizebox{0.15\textwidth}{!}{$ \langle [\mathbf{R}_{\text{cm}}(t)-\mathbf{R}_{\text{cm}}(t_0)]^2 \rangle$} = \frac{6k_B T}{N f_{r}} (t-t_0) + 2\frac{P_b^2|f|^2}{N f_{r}^2} [\tau_{\theta} (t-t_0-\tau_{\theta})  +p\, \tau\, (t-t_0-\tau)]$ in which the diffusion constants due to thermal noise gets modified by its active terms, and interestingly, the contribution not exactly in the same phase as active contribution lags behind with its thermal counterparts. 

\subsection{Calculation of Cooperative diffusivity}
In this section, we study a dilute polyelectrolyte solution comprising a few homogeneously charged flexible polymer chains, each consisting of $N$ monomers, within a volume $V$. The average polymer concentration is $\overline{\rho}_p = n_p/V$, where $n_p$ is the number of polymer chains in the solution. The total charge of the polymer, $\alpha z_p e$, exists before dispersed into the solution, where $e$ is the electric charge, and $\alpha$ is the degree of ionization. Further, the solution contains enzyme-like proteins, also charged, which interact with the polymer. The number of enzymes, $n_{\text{enz}}$, is much smaller than $n_{\text{p}}$, resulting in three species present in the solution: the charged polymer and enzymes, and the dissociated counterions, ensuring overall electrical neutrality, i.e., $\sum^{3}_{i=1} z_{i} e n_i = 0.$ The total number of dissociated counterions is given by $n_c = \left(\alpha \, n_{\text{p}} N \, |\frac{z_{\text{p}}}{z_c}| - n_{\text{enz}} \, |\frac{z_{\text{enz}}}{z_c}|\right)$, where $|z_c \, e|$ denotes the total charge of a single counterion. The dynamics of an enzyme in the solution are driven by thermal forces and screened columbic potentials due to the charged species. While additional forces may arise from chemical reactions or shape of the enzyme, these are beyond the scope of this study. We assume that enzyme binding and unbinding to the polymer are driven by electrostatic and thermal forces, respectively.

\subsection{Derivation of the Fokker-Planck Equation for Polymers}
In this subsection, we focus on deriving the equation for polymer concentration. Let us start with the Stochastic-Liouville equation for polymer density, $\rho(\mathbf{r},t) = \sum_{\alpha} \delta (\mathbf{r}-\mathbf{R}_{\alpha})$, we assume a dilute solution, neglecting interchain entanglement, electrostatic, and hydrodynamic interactions. Incorporating segment-segment interaction and the effects of the friction with the solvents, the large-scale dynamics simplifies to a single charged polymer with active coupling and its dissociated counterions. The time evolution of density is given by
\begin{equation}
\frac{\partial \rho(\mathbf{r},t)}{\partial t} = - \vec{\nabla}\cdot \left[\frac{\partial{\mathbf{R}_{\alpha}}}{\partial t} \, \delta (\mathbf{r}-\mathbf{R}_{\alpha}) \right].
\label{Sto-LiouEqn}
\end{equation}
Hereafter, we drop the chain index $\alpha$ due to the infinitely dilute solution. Let $P(\mathbf{r},t)$ represent the probability density of a polymer, where $\mathbf{r}$ is the position vector of the center of mass. According to the van-Kampen lemma \cite{PhysRep_1976}, $P(\mathbf{r},t) = \langle \, \rho(\mathbf{r},t) \,\rangle$, representing density in phase space. The equation of motion for $q=0$ becomes
\begin{equation}
Nf_r \,\, \frac{\partial }{\partial t} \mathbf{R}_{\text{cm}}(t) = \mathbf{f}_T (t) + \mathbf{f}_{A}(t),
\end{equation}
where we write $\mathbf{f}_T (t)$ and $\mathbf{f}_{A}(t)$ instead of $\widehat{\mathbf{f}}_T (0,t)$ and $\widehat{\mathbf{f}}_{A}(0,t)$ for brevity which are the net thermal and active noises, respectively, experienced by the center of mass of the polymer. Substituting the above equation into \eqref{Sto-LiouEqn} and considering ensemble averages, the time evolution of the probability density is given by
\begin{equation}
\frac{\partial P(\mathbf{r}, t) }{\partial t}  = -\frac{1}{N\,f_r} \vec{\nabla} \cdot \left[ \left\langle (\mathbf{f}_T + \mathbf{f}_A)\, \delta (\mathbf{r}-\mathbf{R}_{\text{cm}})  \right\rangle \right].
\label{ProbFunctional}
\end{equation}
To evaluate the second and third terms on the right-hand side of \eqref{ProbFunctional}, we employ functional calculus \cite{Hanggi1995}. Denoting $F[f_T]$ as a functional of either thermal noise $f_T(t)$ or $f_A(t)$, and $G[f_T]=\delta(\mathbf{r}-\mathbf{R}_{\text{cm}}(t))$, the statistical properties \cite{Hanggi1978} of noise $f_T(t)$ and $f_A(t)$ are used to calculate these terms. Assuming the system is in a stationary state and noise is Gaussian, the correlation between two functionals can be expressed as
\begin{eqnarray}
&&\resizebox{0.25\textwidth}{!}{$\langle F[f_T] G[f_T] \rangle = \langle F[f_T]\rangle \langle G[f_T] \rangle$} + \sum^{\infty}_{n=1} \frac{1}{n!} \medint\int^{t}_{t_0} \dots \medint\int^{t}_{t_0} dt_i \, ds_i \nonumber \\ && \resizebox{0.32\textwidth}{!}{$\left \langle \frac{\delta^n F[f_T]}{\delta f_T(t_1) \dots \delta f_T(t_n)} \right \rangle \left \langle \frac{\delta^n G[f_T]}{\delta f_T(s_1) \dots \delta f_T(s_n)} \right \rangle$} \prod^{n}_{i=1} C(t_i-s_i) 
\label{GenFuncCalMain}
\end{eqnarray}
where $C(t_i-s_i)$ is the two points temporal correlation function. For this work, only the $n=1$ term will contribute, and contributions from all other terms ($n\geq 2$) are zero. Using this equation and \eqref{AppenThermal}, we obtain the correlation for one component of the thermal noise as $\langle f_T(t) \delta (x-R_x(t))\rangle = - k_B T \frac{d}{dx} \langle \delta (x-R_x(t)) \rangle$ wherein $R_x$ is the $x$ component of the position vector of the center of mass.  For colored noise, we obtain (see Appendix for detailed calculation) 
\begin{equation}
\langle f_A \delta(x-R_x) \rangle = \frac{-1}{3 Nf_r} \int^{t}_{t_0} dt' \,\, F_A(t-t')
\frac{d}{d x} \langle \delta(x-R_x) \rangle,
\end{equation}
where $F_A (|t-t'|)$ given in Eq.\,\ref{ActiveForceCorrCM} is the autocorrelation of the active force. Assuming stationary state Gaussian noise, we derive a Fokker-Planck equation as
\begin{eqnarray}
\frac{\partial P(\mathbf{r},t)}{\partial t} &=& \frac{k_B T}{Nf_r} \nabla^2 P + \frac{1}{3 (Nf_r)^2} \medint\int^{\infty}_0 dt' F_A(t') \nabla^2 P  
\label{FKP_final}
\end{eqnarray}
where $P(\mathbf{r},t)$ is the probability density of the center of mass of the polymer. In a dilute solution, the translational diffusion coefficient in the equation above follows Zimm dynamics\,\cite{Kirkwood_1948,Zimm1956,Doi1988}, dominated by hydrodynamic interactions. In terms of polymer concentration, this simplifies to
\begin{eqnarray}
\frac{\partial \rho_{p}(\mathbf{r},t)}{\partial t} &=& D_{\text{eff}} \,\nabla^2 \rho_{p} (\mathbf{r},t) \quad
\label{Con_Final}
\end{eqnarray}
where $D_{\text{eff}} = \bigl[k_B T/(N f_r) + (\tau_{\theta}+ p\,\tau) \,  |f|^2 P^2_b/(3 N\,f^2_r) \bigr].$ In the absence of excluded volume and electrostatic interactions, $1/f_r = 1/\zeta + 8 \sqrt{2}\,N/(3 (12 \pi^3 L \ell)^{1/2} \eta_0).$ Incorporating screened excluded volume and electrostatic interactions, the Kuhn length $\ell$ is modified to $\ell_1$, yielding $ 1/f_r = 1/\zeta + 8 \sqrt{2}\, N/(3 (12 \pi^3 N\ell \ell_1)^{1/2} \eta_0).$ In the free-draining limit, the diffusion constant of polymer is given by $D_p = k_B T/\zeta N$. When hydrodynamic interaction dominates,
\begin{eqnarray}
D_p = \frac{8 \sqrt{2}}{3} \, \frac{k_B T }{\eta_0 \left(12 \pi^3 L\ell_1\right)^{1/2} }. 
\end{eqnarray}
The contribution of active coupling of enzymes on the diffusion constant of polymer is obtained  as 
\begin{eqnarray}
D_A &=& (\tau_{\theta}+ p\,\tau) \,  \frac{|f|^2 P^2_b}{3 N\,f^2_r}.
\label{CompactActiveDiff} 
\end{eqnarray}
This shows how the contribution to diffusion constant due to the active force $D_A$ explicitly depends on various factors such as $N$, $\tau$ and effective Kuhn length $\ell_1$. The expression for $\ell_1$ \cite{JCP_Muthu1982,Muthukumar1987}, capturing the effects of electrostatic interactions between segments and excluded volume, is given by $ \ell^{3/2}_1 (\ell^{-1} - \ell^{-1}_1) = \zeta_{\text{ev}} + \zeta_{\text{charge}},$ where 
$ \zeta_{\text{ev}} = (4/3)\,  (3/2\pi)^{3/2} (w L^{1/2})/\ell_1 $ and $ \zeta_{\text{charge}} = (4/45)\, (6/\pi)^{1/2} w_c L^{3/2} [\frac{15 \pi^{1/2} e^a}{2 a^{5/2}} (a^2-4a+6) \text{Erfc}(\sqrt{a})-\frac{3\pi}{a^{5/2}} + \frac{\pi}{a^{3/2}} + \frac{6 \sqrt{\pi}}{a^2} ] $
with $a= (\kappa^2 L \ell_1)/6$.
\paragraph{Degree of Activity:} We define the strength of activity as the ratio of the contribution to diffusivity from active to thermal parts. In this work, we define the Péclet number, $\text{Pe}= D_A/D_p$ as
\begin{eqnarray}
\text{Pe} &=& (\tau_{\theta}+p\, \tau)\, \,\frac{P_{b}^2 \, |f|^2 }{3 k_B T} \left(\frac{1}{\zeta} + \frac{8 \sqrt{2}N }{3 (12 \pi^3 N\ell \ell_1)^{1/2} \eta_0}\right),
\end{eqnarray}
which is dimensionless and indicates the strength of activity.


\subsection{Counterion Coupling}
Here we consider the coupled dynamics of counterions and derive the cooperative diffusivity of the polymer in an infinitely dilute solution. As a charged polymer in solution is always surrounded by its dissociated counterion clouds, the dynamics of counterions and polymer are coupled and significantly modified by this interaction. Though the overall solution is electrically neutral, the presence of counterions leads to density fluctuations and induces a local electric field $\mathbf{E}(\mathbf{r}, t)$, which affects the polymer dynamics. In the absence of active coupling, when the charged polymeric solution is infinitely dilute, the scenario simplifies and can be described by the well-known Nernst-Hartley theory \cite{Berne1976}. This leads Eq. \ref{FKP_final} to become an equation of continuity:
\begin{equation}
\frac{\partial \rho_p(\mathbf{r},t)}{\partial t} = - \bf \nabla \cdot \mathbf{J}_p  \,\,\,\, \text{and} \,\,\,\,  \mathbf{J}_p = - \, \nabla \cdot [D_{\text{eff}} \, \rho_p ] + \rho_p  \mu_p \mathbf{E}  
\label{ContinuityPolymer}
\end{equation}
where $\rho_p(\mathbf{r},t)$ is the local polymer concentration. The dynamics of the concentration of counterions are strongly coupled to the dynamics of the polymer concentration. The equation of motion for counterions can be written as:
\begin{equation}
\frac{\partial \rho_c(\mathbf{r},t)}{\partial t} = - \bf \nabla \cdot \mathbf{J}_c  \,\,\,\, \text{and}  \,\,\,\, \mathbf{J}_c = - \, \nabla \cdot [D_c \, \rho_c ] +  \rho_c \mu_c \mathbf{E} 
\label{Current_EqContinuity1}
\end{equation}
where $\rho_c(\mathbf{r},t)$ is the local counterion concentration. These three species are dynamically different, and since time-dependent concentration fluctuations of the species can be measured in dynamic light scattering experiments, we express concentration fluctuations up to linear order as $\rho_p(\mathbf{r},t) = \overline{\rho}_p + \delta \rho_p (\mathbf{r},t)$, $\rho_{c}(\mathbf{r},t) = \overline{\rho}_c + \delta \rho_c (\mathbf{r},t)$, and $\rho_{\text{enz}} (\mathbf{r},t) = \overline{\rho}_{\text{enz}} + \delta \rho_{\text{enz}} (\mathbf{r},t)$, where $\overline{\rho}_p$, $\overline{\rho}_c$, and $\overline{\rho}_{\text{enz}}$ are the average concentrations of polymer, counterions, and charged enzymes, respectively. The divergence of the electric field can be written as $
\nabla \cdot \mathbf{E} = \frac{4 \pi e}{\epsilon} \left[ \alpha \, z_p \,N\rho_p  + z_{\text{enz}} \, \rho_{\text{enz}} + z_c \, \rho_c \right] $ where $\alpha$ is the degree of ionization, and each segment of a polymer chain contains charge $\alpha z_p e$. The enzyme density is considered to be very low, $\rho_{enz} \ll \rho_c$. To study the coupled dynamics, we define the Fourier transform of $\delta \rho_p(\mathbf{r},t)$ as $\delta \rho_i(\mathbf{r}, t) = \int \frac{d^3 \mathbf{k}}{(2\pi)^3} \,\, \delta \rho_i(\mathbf{k},t) \,\, e^{i \mathbf{k} \cdot \mathbf{r}}$, where $\mathbf{k}$ is the scattering wave vector. The time evolution equation of the small fluctuations in the concentration field of the polymer and counterions, up to linear order, is obtained \cite{MuthuACP2005}. Using electroneutrality, we have the coupled equations in Fourier space:
\begin{equation}
  \begin{split}
\frac{\partial }{\partial t} \delta \rho_p (\mathbf{k},t) = - D_{\text{eff}} \, \mathbf{k}^2 \, \delta \rho_p  -\frac{4 \pi}{\epsilon} \mu_p   \overline{\rho}_p e \, \bigl[ \alpha z_p \,N \delta \rho_p    + z_c \, \delta \rho_c \bigr] 
\end{split}
\end{equation}

and 
\begin{equation}
  \begin{split}
\frac{\partial }{\partial t} \delta \rho_c(\mathbf{k},t) = - D_{c} \, \mathbf{k}^2 \delta \rho_c  -\frac{4 \pi}{\epsilon} \mu_c  \overline{\rho}_c e \bigl[ \alpha z_p \, N \delta \rho_p  + z_c \delta \rho_c \bigr] 
\label{ConFlucCounter}
\end{split}
\end{equation}
where $\epsilon$ is the dielectric constant of the solution $\mu_c=e z_c D_c/k_B T$, and $\mu_p=\alpha z_p\,N e D_p/k_B T$ (at high counterion concentrations). The relaxation time scale of the counterions is much faster compared to macromolecules. Therefore, the fluctuation of polymer concentration experiences the steady-state nature of the counterions, i.e., $\frac{\partial }{\partial t} \delta \rho_c(\mathbf{k},t) =0$. Considering the expression of $\delta \rho_c$ in steady state, we obtain the expression for $\delta \rho_p$ as
\begin{equation}
\frac{\partial }{\partial t} \delta \rho_p(\mathbf{k},t) = - D_{\text{coop}} \,\, k^2 \, \delta \rho_p (\mathbf{k},t)  
\label{DensityFlucCoupledFinal}
\end{equation}
where $D_{\text{coop}} = D_{\text{eff}} + 4\pi/\epsilon\, (\mu_p (\alpha z_p e)\, N \overline{\rho}_p)/(\mathbf{k}^2+\kappa^2_c)$ and $\kappa^2_c =(4\pi/\epsilon)\,(z_c e)^2\, \overline{\rho}_c/k_B T$. We also have $ D_{\text{eff}} = D_p \left( 1+ \text{Pe}\,\right)$. We consider $\mathbf{k} \rightarrow 0$, and substitute $\mu_p= \alpha z_p e\,N D_p/k_B T$  and $\kappa^2_c$ in the expression for $D_{\text{coop}}$, and get
\begin{equation}
D_{\text{coop}}\, = \,D_p \left(1 +\text{Pe} + \alpha \frac{z_p}{z_c} N \right)
\end{equation}
where we use the electroneutrality condition of the polymer and its counterions $\alpha z_p N \overline{\rho}_p= z_c \overline{\rho}_c$. 
The above equation indicates that for a fixed $\alpha$, the cooperative diffusivity increases linearly with the degree of activity $\text{Pe}$ and $N$. Solving Eq. (\ref{DensityFlucCoupledFinal}), we obtain the time-dependent polymer density-density fluctuations as:
\begin{equation}
\langle \delta \rho_p(\mathbf{k},t) \delta \rho_p(\mathbf{k},0) \rangle \sim e^{-D_{\text{coop}}\, k^2 \, t}. \label{DenFluc}
\end{equation}
This equation is important because the cooperative diffusivity can be measured in dynamic light scattering, where $k$ is the scattering wave vector.

\section{Discussion and Conclusions}

Motivated by real-life biological and synthetic bio-polymeric systems, we investigate the transport properties of a charged macromolecular system. In this system, enzymes bind and unbind to the macromolecule, affecting its transport properties. Initially, we study equal-time segment-segment correlations and mean square displacement (MSD), considering electrostatic binding as the source of active coupling. We express our findings in terms of the absolute value of force ($|F|$) which is a general parameter. This approach allows us to study various types of local forces that may cause colored noise. Furthermore, we present our results in terms of the size exponent ($\nu$), making them applicable for any feasible value of $\nu$. Next, we focus on a dilute solution where interactions between macromolecules and competitive or cooperative effects of enzymes are negligible. We study how the transport properties, particularly the diffusion constant of the macromolecule, are modified by the active coupling of enzymes and how these changes depend on the environment of the macromolecule.

To address this question, we investigate the scenario analytically at two levels. First, we study the modified dynamics of individual segments and the static structure of the macromolecule, revealing a new regime dominated by active coupling that lead to macromolecule swelling. Second, we study the concentrations of macromolecules, observing their coupling with dissociated counterions, which is experimentally measurable. Our obtained closed-form expression for activity-induced cooperative diffusivity depends on parameters such as degree of polymerization, degree of ionization, temperature, counterion concentration, and binding and unbinding properties of enzymes, as well as other transport properties. 

Nevertheless, the expression for diffusivity does not account for dynamics of enzyme such as  shape and conformational changes during binding and unbinding. Though the details of the dynamics of enzyme may be crucial in a real-life biological system, namely cell, the environment is crowded and salty which are beyond the scope of this work. Even in the absence of active coupling, the dynamics of a charged macromolecule in a simple polyelectrolyte solution are significantly affected by the concentration of added salt. This is due to the regulation of electrostatic screening, which leads to notable changes in dynamics. Considering active coupling, investigating the effect of added salt on relaxation modes will be an important future study, both in dilute and high concentration limits.

\section*{ACKNOWLEDGMENT}
 Acknowledgement is made to the National Institutes of Health (Grant No. 5R01HG002776-16), the National Science Foundation (DMR-2015935), and the AFOSR Grant FA9550-20-1-0142 for financial support. TS acknowledges funding support from INSERM ITMO Cancer (Grant Number: 20CR110-00).

\appendix
\section{Calculation of the autocorrelation}
\subsection{Master equation for bound and unbound state}
Let us consider that $A'$ is the initial state at time $t'$, in which the possible states of $A'$ would be either bound ($+$) or unbound ($-$). The initial state $A'$ evolves to the new states $+$ (bound) with transition probability $P(+,t|A', t')$ and $-$ (unbound) with $P(-,t|A', t')$. The evolution of the initial state $A'$ can be described as
\begin{equation}
\begin{split}
\frac{\partial P(+,t|A, t')}{\partial t} = -\lambda_{+} \, P(+,t|A', t') + \lambda_{-} \, P(-,t|A', t') \\ 
\frac{\partial P(-,t|A', t')}{\partial t} = -\lambda_{-} \, P(-,t|A', t') +\lambda_{+} \, P(+,t|A', t')  
\label{Two-StateEqns}
\end{split}
\end{equation}
The steady state probabilities of bound and unbound states can be obtained as 
$ P_b = \lambda_{-}/\bigl(\lambda_{+}+\lambda_{-}\bigr) \, ,  P_{ub} =\lambda_{+}/\bigl(\lambda_{+}+\lambda_{-}\bigr) $ we solve the Eq.\,\ref{Two-StateEqns}, and obtain 
\begin{equation}
\begin{split}
 P(+,t|A',t') = P_b  +  e^{-(\lambda_{+}+\lambda_{-})|t-t'|} \bigl(P_{\text{ub}} \delta_{+,A'} - P_b \delta_{-,A'} \bigr) \\
P(-,t|A',t') = P_{\text{ub}} + e^{-(\lambda_{+}+\lambda_{-})|t-t'|} \bigl(P_{\text{b}} \delta_{-,A'} - P_{\text{ub}} \delta_{+,A'} \bigr)  \,
\end{split}
\label{TransProb}
\end{equation}

\subsubsection{Auto-correlation of telegraphic forces}
We consider that initially, the system was in the $A'$ state at $t'$.  This indicates, in general, that a few or many bound and unbound events might have occurred in this time span ($|t-t'|$). Therefore, in principle, the orientation of an active force $\mathbf{f}_A (t)$ is different from that of $\mathbf{f}_{A'} (t')$. As we mentioned before, the state $A$ consists of $+$ and $-$, the force-force correlation can be written as 
\begin{align}
\nonumber
    \langle \mathbf{f}_{A}(t) \cdot \mathbf{f}_{A'}(t') \rangle =& \sum_{A=+,-} \sum_{A'=+,-} \langle \mathbf{f}_A(t)|A',t\rangle   \cdot \mathbf{f}_{A'}(t') P_s(A') \\
\nonumber        
    =&\,\,\langle \mathbf{f}_+(t)|+,t\rangle   \cdot \mathbf{f}_{+}(t') P_s(+),
\label{Corr}
\end{align}
where we have used the fact that  $\mathbf{f_-=0}$ that will not contribute to the auto-correlation. The magnitude of binding force  $|\mathbf{f}|=f_+$ is the same at $t$ and $t'$, however, the direction of $\mathbf{f}(t)$ is different from $\mathbf{f}(t')$ generally, though they are correlated. Therefore, using the transition probability from Eq.\,\ref{TransProb}  in Eq.\ref{Corr}, we obtain
\begin{align}
    \langle \mathbf{f}_{A}(t) \cdot \mathbf{f}_{A'}(t') \rangle=&f_+^2\,P(+,t|+,t')\,P_s(+)  \langle \hat{n}(t)|\hat{n}'(t')\rangle  \cdot \hat{n}'(t')
\end{align}
which can be written as
\begin{equation}
  \begin{split}
\langle \mathbf{f}_{A}(t) \cdot \mathbf{f}_{A'}(t') \rangle    =f^2_+ \bigl[(\frac{\lambda_{-}}{\lambda_{+}+\lambda_{-}})^2 +\frac{\lambda_{+}\lambda_{-}}{(\lambda_{+}+\lambda_{-})^2}  \\ e^{-(\lambda_{+}+\lambda_{-})|t-t'|}\bigr]  \times \langle \cos \theta (|t-t'|)  \rangle,  
\end{split}
\end{equation}
where we have substituted the form of $P(+,t|+,t')$ and $\langle \hat{n}(t)|\hat{n}'(t')\rangle$ means that given the direction $\hat{n}(t')$ of $\mathbf{f}_+(t')$, the expectation value of $\hat{n}(t)$ of $\mathbf{f}_+(t)$, and the inner product is $\langle \cos \theta (|t-t'|)  \rangle$ .

\subsubsection{Calculation of $\langle \cos \theta(|t-t'|) \rangle$}
We assume the diffusive equation,
\begin{align}
    \frac{\partial P(\theta, \phi,t)}{\partial t} =\frac{1}{2\tau_{\theta}}\left(\frac{1}{\sin \theta}\frac{\partial }{\partial \theta}\sin \theta \frac{\partial }{\partial \theta}+\frac{1}{\sin^2\theta}\frac{\partial^2}{\partial \phi^2}\right)P 
    \label{ProbDiffuEq}
\end{align}
The initial distribution can be obtained as 
\begin{align}
    P(\theta,\phi,t=0)=\delta\left(\cos\theta-1\right)\delta\left(\phi\right)
    \label{iniCond}
\end{align}
which satisfies $ \int d \Omega \,\, P(\theta,\phi,t=0) = \int^{2\pi}_{0} d\phi \int^{\pi}_{0} d\theta \, \sin (\theta) \, \delta\left(\cos\theta-1\right)\delta\left(\phi\right) = 1$ (the normalization condition) where $d\Omega= \sin{\theta} \,d\theta \,d\phi$. 
The solution to Eq.\,(\ref{ProbDiffuEq}) can be obtained as 
\begin{align}
    P(\theta,\phi,t)=\sum^{\infty}_{\ell=0}\sum_{m=-\ell}^{\ell}A_{\ell}^m \, e^{-\frac{\ell(\ell+1)t}{2\tau_R}} \,\, Y^m_\ell(\theta,\phi).
    \label{SolDiffEq}
\end{align}
where $Y^m_\ell(\theta,\phi)$ the spherical harmonics can be expressed as
\begin{align}
    Y^m_\ell(\theta,\phi)=\sqrt{\frac{(2\ell+1)}{4\pi}\frac{(\ell-m)!}{(\ell+m)!}}P^m_\ell(\cos\theta)e^{ i m\phi},
\end{align}
are the complete and orthogonal basis for the Hilbert space $\theta\in(0,\pi) $, $\phi \in (0,2\pi)$ 
with the inner product leads $
\int^{\pi}_{0} d\theta \sin (\theta) \int^{2\pi}_{0}  d\phi \,\,  Y^{m}_{\ell} (\theta, \phi) \,\,  Y^{m'*}_{\ell'} (\theta, \phi) = \delta_{\ell,\ell'} \,\, \delta_{m,m'}$. We multiply $Y^{m'*}_{\ell'}(\theta, \phi)$, where $*$ means the complex conjugate, with  Eq.\,(\ref{SolDiffEq}), and integrate over solid angle $d\Omega$ of the above equation at time $t=0$,  we obtain 
\begin{equation}
Y^{m'*}_{\ell'}(\cos^{-1}(1),0) =  \sum^{\infty}_{\ell=0}\sum_{m=-\ell}^{\ell} A^{m}_{\ell} \,\, \delta_{\ell,\ell'} \,\, \delta_{m,m'} 
\end{equation}
where we use the initial distribution given by Eq.\,(\ref{iniCond}). This leads to 
\begin{equation}
 A^{m}_{\ell} = Y^{m*}_{\ell}(\cos^{-1}(1),0) .\label{A}
\end{equation}
Using the probability distribution of $\theta$ and $\phi$, we can write 
$ \langle \cos \,(\theta ) \rangle=\int  d\Omega \,\, P(\theta,\phi,t) \cos(\theta).$ Substituting the form of $P(\theta, \phi,t)$ from 
Eq.\,(\ref{SolDiffEq}) and $A^{m}_{\ell}$ from Eq.\,\ref{A}, we obtain
\begin{eqnarray}
\langle \cos \,(\theta ) \rangle= \sum_{\ell=0}\sum_{m=-\ell}^{\ell}  e^{-\frac{\ell(\ell+1)t}{2\tau_{\theta}}} \,\, && \medint\int^{\pi}_{0} d\theta \sin(\theta) \medint\int^{2\pi}_{0} d\phi \,\,  \cos(\theta) \,\, \nonumber \\ && Y^{m*}_\ell(0,0) \,\, Y^m_\ell(\theta,\phi)
\end{eqnarray}
In the above equation, $\ell =1$ and $m=0$ would only contribute, other combination of $\ell$ and $m$ would give vanishing result. We substitute the form $Y^{0}_{1}(0,0)= (1/2) \sqrt{3/\pi}$ and $Y^{0}_{1}(\theta,\phi)=(1/2) \sqrt{3/\pi} \,\, \cos{\theta}$ in the above equation and obtain
\begin{eqnarray}
\langle \cos \,(\theta ) \rangle =e^{-t/\tau_{\theta}}
\end{eqnarray}

\section{Segment-to-Segment Correlation function}
We study the segment-to-segment correlation and MSD of a segment with the active coupling. We write segment-to-segment correlation function in Fourier-space as  
\begin{eqnarray}
&& \resizebox{0.14\textwidth}{!}{$\langle [\mathbf{R}(s,t)-\mathbf{R}(s',t')]^2  \rangle$} = \medint\int\medint\int\medint \int\medint\int\frac{dq\,dq'd\omega\,d \omega'}{(2\pi)^4}  \resizebox{0.14\textwidth}{!}{$\langle \mathbf{R}(q,\omega) \cdot \mathbf{R}(q',\omega')  \rangle$} \nonumber \\
&& \times [ e^{i(q+q')s+i(\omega+\omega')t} + e^{i(q+q')s'+i(\omega + \omega')t'}  
- 2\, e^{i(q s + q' s')+i(\omega t+ \omega' t')}] \nonumber \\
\label{EqStrucFunc}
\end{eqnarray}

Expressing the correlation in $q$ and $\omega$ space, we carry out the frequency integrations, let us present the segment-segment correlation in Fourier space  as
\begin{eqnarray}
\resizebox{0.14\textwidth}{!}{$\langle \R (q,\omega) \cdot \R(q',\omega') \rangle$} &=& H^2(q) \left(\resizebox{0.2\textwidth}{!}{$\frac{\langle \hat{\mathbf{f}}_{T}(q,\omega) \hat{\mathbf{f}}_{T}(q', \omega')\rangle}{(i \omega + q^2/\tilde{\tau}_1(q))(i\omega' +q'^2/\tilde{\tau}_1(q))}$}  \right. \nonumber\\ && \left.+  \resizebox{0.2\textwidth}{!}{$\frac{\langle \hat{\mathbf{f}}_{A}(q,\omega) \hat{\mathbf{f}}_{A}(q', \omega')\rangle}{(i \omega + q^2/\tilde{\tau}_1(q))(i\omega' +q'^2/\tilde{\tau}_1(q))}$} \right)
\label{FT_SegmentSegment}
\end{eqnarray}
It is important to mention that here we are interested in studying the effect of all modes except the motion of the center of mass ($q=0$). Therefore, in order to express the above equation strictly for $q\neq 0$, we use
\begin{equation}
\begin{split}
\resizebox{0.47\textwidth}{!}{$\R(s,t) = \widehat{\R}(0, t) + \int^{\infty}_{-\infty} \frac{dq}{[2\pi]} \,  e^{i q s} \, \, \widehat{\R}(q,t) -  \int^{q_{\text{low}}}_{-q_{\text{low}}} \frac{dq}{[2\pi]}    e^{i q s} \, \, \widehat{\R}(q,t)$}
\label{FT}
\end{split}
\end{equation}
Performing $\omega'$ and $q'$ integrations, and  we have 
\begin{eqnarray}
\langle[\mathbf{R}(s,t)&-&\mathbf{R}(s',t')]^2\rangle = 2 \medint\int^{\infty}_{-\infty} \frac{dq}{(2\pi)}\medint\int^{\infty}_{-\infty}\frac{d\omega}{(2\pi)} 
\left( 6 k_B T H(q)  \right. \nonumber \\ && \left. +2 |f|^2\, P^2_b \, H^{2}(q)  (\frac{\tau_{\theta}}{1+\omega^2/\tau_{\theta}^2}+ \frac{p\, \tau}{1+\omega^2/\tau^2} ) \right) \nonumber\\ && 
\times \frac{[1-\cos(|s-s'| q)e^{-i\omega (t-t')}]}{\omega^2+ (\frac{q^2}{\tilde{\tau}_1(q)})^2} - \Delta (q_{\text{low}})
\label{MonomerDisplaceWhole}
\end{eqnarray}
where 
\begin{eqnarray}
&&\Delta (q_{\text{low}}) = \Delta_{\text{T}} + \Delta^{(1)}_{\text{A}} +\Delta^{(2)}_{\text{A}} \nonumber \\ && = 2 \, \int^{q_{\text{low}}}_{-q_{\text{low}}}\frac{dq}{(2\pi)}\,\int^{\infty}_{-\infty} \frac{d\omega}{(2\pi)}\,  [6 k_B T H(q)  +2 |f|^2\, P^2_b \, H^{2}(q) \nonumber \\ && (\frac{\tau_{\theta}}{1+(\omega\tau_{\theta})^2}+ \frac{p\, \tau}{1+(\omega\tau)^2} ) ]  \frac{[1-\cos(|s-s'| q)e^{-i\omega (t-t')}]}{\omega^2+ \left(\frac{q^2}{\tilde{\tau}_1(q)}\right)^2} \nonumber \\
\end{eqnarray}
We shall also split up the correlation into three parts, namely thermal and active for the sake of presentation as $ \langle [\R(s,t)-\R(s',t')]^2  \rangle = \langle [\R(s,t)-\R(s',t')]^2  \rangle_{\text{T}} +\langle [\R(s,t)-\R(s',t')]^2  \rangle^{(1)}_{\text{A}} +\langle [\R(s,t)-\R(s',t')]^2  \rangle^{(2)}_{\text{A}} .$
We first carry out the frequency integrations for both the parts and obtain
\begin{equation}
  \begin{split}
\langle[\R(s,t)-\R(s',t')]^2\rangle_{\text{T}} = 6\,k_B T \int^{\infty}_{-\infty} \frac{dq}{(2\pi)}  H(q) \frac{\tilde{\tau}_1(q)}{q^2} \\ \times \left(1-\cos(|s-s'| q)e^{-\frac{|t-t'|}{\tilde{\tau}_1(q)} q^2}\right)-\Delta_{\text{T}} (q_{\text{low}})
\label{FTGenCorrThermal}
\end{split}
\end{equation}
The contribution to the active part can be obtained as
\begin{equation}
\begin{split}
\langle[\R(s,t)-\R(s',t')]^2\rangle^{(1)}_{\text{A}}  = 2\, \, \tau_{\theta} \, P^2_{b} |f|^2  \int^{\infty}_{-\infty} \frac{dq}{[2\pi]}\,\frac{H^2(q)}{\left[1-\bigl(\frac{\tau_{\theta}}{\tilde{\tau}_1(q)} q^2\right)^2 \bigr]} \\
\bigl[\frac{\tilde{\tau}_1(q)}{q^2}\bigl( 1-\cos( q|s-s'|)e^{-\frac{|t-t'|}{\tilde{\tau}_1(q)} q^2}\bigr) - \\ \tau_{\theta} \, \bigl(1-\cos(q|s-s'|)e^{-\frac{|t-t'|}{\tau_{\theta}} }\bigr)\bigr]  - \Delta_1 (q_{\text{low}})
 \label{FTGenCorrActive1st}
\end{split}
\end{equation}
The second part becomes
\begin{equation}
\begin{split}
\langle[\R(s,t)-\R(s',t')]^2\rangle^{(2)}_{\text{A}} = 2\, p\, \tau \, P^2_b\,  |f|^2 \int^{\infty}_{-\infty} \frac{dq}{[2\pi]}\,\frac{H^2(q)}{[1-\bigl(q^2\tau/\tilde{\tau}_1(q) \bigr)^2]}\\
 \bigl[ \frac{\tau_1(q)}{q^2}\bigl( 1-\cos( q|s-s'|)e^{-\frac{|t-t'|}{\tilde{\tau}_1(q)} q^2}\bigr)  -\tau\bigl(1-\cos(q|s-s'|) \\ e^{-\frac{|t-t'|}{\tau} }\bigr)\bigr] - \Delta_2 (q_{\text{low}})
\label{FTGenCorrActive2nd}
\end{split}
\end{equation}


\subsection{Time independent segment-segment correlation}

\subsubsection{Thermal contribution}
We substitute the form of $H(q)$ and $\tilde{\tau}_1(q)$, and obtain the thermal contribution at $t=t'$  as 
\begin{equation}
\begin{split}
\langle[\R(s,t)-\R(s',t)]^2\rangle_{\text{T}} & = 2\, a_{\ell} \, \ell^2 \int^{\infty}_{-\infty} \frac{dq}{(2\pi)} \, |q|^{-1-2\nu} \left(1- \right. \\ 
& \left. \cos(|s-s'| q)\right) -\Delta_{\text{T}}(q_{\text{low}}) \\
& \simeq 2\,\frac{a_{\ell}}{\pi}\,\ell^2 \cos(\pi (1+\nu)) \, \Gamma(-2\nu) \, |s-s'|^{2\nu}
\end{split}
\end{equation}
the second term ($\Delta_{\text{T}}(q_{\text{low}})$) can be obtained as $\Delta_{\text{T}}(q_{\text{low}}) \sim q^{2(1-\nu)}_{\text{low}}$. For the size exponent $\nu < 1$ as $N \rightarrow \infty$, the contribution from the mode $q_{\text{low}}$ becomes very small and consequently, $\Delta_{\text{T}}(q_{\text{low}}) $ is negligible compared with the first term.   In  the above expression, the information of the electrostatic and hydrodynamic interactions are buried in the parameter $a_{\ell}$. 
\subsubsection{Active force contribution}
At $t=t'$, substituting $H(q)$ and $\tilde{\tau}_1(q)$ in Eq.\,(\ref{FTGenCorrActive1st}), we have 
\begin{equation}
\begin{split}
\langle[\R(s,t)-\R(s',t)]^2\rangle^{(1)}_{\text{A}}  = 2\, \tau_{\theta} \, P^2_{b}  |f|^2  \int^{\infty}_{-\infty} \frac{dq}{[2\pi]}\, \\ \frac{H^2(q)}{\bigl[1+(\frac{\tau_{\theta}}{\tilde{\tau}_1(q)} q^2) \bigr]}
 \frac{\tilde{\tau}_1(q)}{q^2}(1-\cos(q|s-s'|)) -\Delta_1 (q_{\text{low}})
\end{split} 
\end{equation}
which can be obtained as 
\begin{equation}
\begin{split}
\langle[\R(s,t)-\R(s',t)]^2\rangle^{(1)}_{\text{A}}  = 2 \tau_{\theta} P^2_b \,|f|^2  \frac{a^2_h }{\pi \zeta^2}\, \overline{\tau}_1 \, |s-s'|^{1+\nu} \\ J(u_{\tau}) -\Delta_1 (q_{\text{low}})
\end{split} 
\end{equation}
and similarly from Eq.\,(\ref{FTGenCorrActive2nd}), we have
\begin{equation}
\begin{split}
\langle[\R(s,t)-\R(s',t)]^2\rangle^{(2)}_{\text{A}} = 2\,p \, \tau \,  |f|^2 P^2_{b} \int^{\infty}_{-\infty} \frac{dq}{[2\pi]}\, \\ \frac{H^2(q)}{\left[1+(\frac{\tau}{\tilde{\tau}_1(q)} q^2) \right]}
 \frac{\tilde{\tau}_1(q)}{q^2}(1-\cos(q|s-s'|))-\Delta_2 (q_{\text{low}})
 \end{split}
\end{equation}
similarly, we obtain
\begin{equation}
\begin{split} 
\langle[\R(s,t)-\R(s',t)]^2\rangle^{(2)}_{\text{A}} = 2 p\, \tau P^2_b |f|^2  \frac{a^2_h \,}{\pi \zeta^2}\,  \overline{\tau}_1 |s-s'|^{1+\nu} \\ \times J(v_{\tau}) -\Delta_2 (q_{\text{low}})
 \end{split}
\end{equation}
We also substitute the expression for $\tilde{\tau}_1(q)$ and $H(q)$  in the above equation, and obtain $\langle [\R(s,t)-\R(s',t)]^2 \rangle_{\text{A}} = \langle [\R(s,t)-\R(s',t)]^2 \rangle^{(1)}_{\text{A}} +\langle [\R(s,t)-\R(s',t)]^2 \rangle^{(2)}_{\text{A}}$
\begin{equation}
\begin{split}
\langle [\R(s,t)-\R(s',t)]^2 \rangle_{\text{A}}  \approx 2 \,P^2_b  |f|^2 \,\frac{a^2_{h}}{\pi \zeta^2}\, \overline{\tau}_1 \bigl( \tau_{\theta} \, J(u_{\tau}) \nonumber \\ + p \, \tau \, J(v_{\tau})\bigr)   |s-s'|^{1+\nu}  -\Delta_1 (q_{\text{low}})-\Delta_2 (q_{\text{low}})
\end{split}
\end{equation}
where 
\begin{equation}
J(u_{\tau})=\int_{0}^{\infty}du\,\frac{1-\cos u}{ u^{\nu+2}}\frac{1}{1+(\frac{u}{u_{\tau}})^{3\nu}}
\label{EqGAZIntegral}
\end{equation}
and $u_{\tau}^{3\nu} = \frac{\tau_1}{\tau_{\theta}}\frac{a_{\ell}}{a_{h}}|s-s'|^{3\nu} = \frac{\tilde{\tau}_1}{\tau_{\theta}} |s-s'|^{3\nu}$, and $v_{\tau}^{3\nu} = \frac{\tau_1}{\tau}\frac{a_{\ell}}{a_{h}}|s-s'|^{3\nu} = \frac{\tilde{\tau}_1}{\tau} |s-s'|^{3\nu}$. 
\begin{align}
J(u_{\tau}) =   \begin{cases}
          \frac{\Gamma(\frac{1}{3\nu}-\frac{1}{3})}{6\nu}  \Gamma(\frac{4}{3}-\frac{1}{3\nu})\, u_{\tau}^{1-\nu} &\text{for $u_{\tau} \ll 1 $}  \\
          \frac{\sin(\frac{\pi \nu}{2})}{\pi (1+\nu)}  \Gamma(\frac{(1-\nu)}{2}) \,\,  \quad &\text{for $u_{\tau} \gg 1 $ } 
\end{cases}
\end{align}
The form of expression will be similar for $J(u_{\tau}).$ As $1/\tau = 1/\tau_{\text{b}}+1/\tau_{\text{ub}}+1/\tau_{\theta}$, therefore, $\tau < \tau_{\theta}$. Depending on the these condition there are three regimes. First, $|s-s'| \ll (\tau/\overline{\tau}_1)^{1/3\nu}$, secondly, $|s-s'| \gg (\tau_{\theta}/\overline{\tau}_1)^{1/3\nu}$ and third regime is $(\tau/\overline{\tau}_1)^{1/3\nu} \ll |s-s'| \ll (\tau_{\theta}/\overline{\tau}_1)^{1/3\nu}$. 
  For finite $\tau$ and $\tau_{\theta}$, when the degree of polymerization of the polymer is large i.e., $N$ is very large, $q_{\text{low}}$, 
    becomes very small, and the value of $\Delta_1 (q_{\text{low}})$ and $\Delta_2(q_{\text{low}})$ becomes $\sim q^{1-\nu}_{\text{low}}$. 
    Therefore, compared with the other term the contribution of the  terms ($\Delta_1$, and $\Delta_2$) become  negligibly small.  However, we have a few different regimes.
\subsubsection{Different regimes}  

\paragraph{\bf First regime}
Performing the integration for the active part and for $t=t'$, we consider the scaling limit $|s-s'| \ll \bigl(\tau/\overline{\tau}_1\bigr)^{1/3\nu} $, and obtain
\begin{equation}
\begin{split}
\langle [\R(s,t)-\R(s',t)]^2 \rangle_{\text{A}} \simeq 2\, \frac{b(\nu)}{\pi}\,\, \left(\frac{P_b |f|\, a_h \overline{\tau}_1}{\zeta}\right)^2\,  \bigl(\frac{\tau_{\theta}}{\overline{\tau}_1}\bigr)^{\frac{4}{3}-\frac{1}{3\nu}} \\ (1+p \left(\tau/\tau_{\theta}\right)^{\frac{4}{3}-\frac{1}{3\nu}}) \,  |s-s'|^2 
\end{split}
\end{equation} 
where $b(\nu)=\frac{1}{6\nu} \Gamma(\frac{1}{3\nu}-\frac{1}{3}) \Gamma(\frac{4}{3}-\frac{1}{3\nu}).$

\paragraph{\bf Second regime}
Similarly, in the limit $|s-s'| \gg \bigl(\tau_{\theta}/(\overline{\tau}_1)\bigr)^{1/3\nu} $, we obtain
\begin{equation}
\begin{split}
\langle [\R(s,t)-\R(s',t)]^2 \rangle_{\text{A}} \simeq
2\,\, \frac{g(\nu)}{\pi}\,\, \left(\frac{P_b |f|\, a_h\, \overline{\tau}_1}{\zeta}\right)^2  \,\left(\frac{\tau_{\theta}}{\overline{\tau}_1}\right) \\ \, (1+p\, (\frac{\tau}{\tau_{\theta}}) ) \,\,  |s-s'|^{1+\nu}  
\end{split}
\end{equation}
where $g(\nu)=\frac{1}{\pi (1+\nu)} \sin(\frac{\pi \nu}{2}) \Gamma(\frac{(1-\nu)}{2}).$ There is a crossover regime where  $(\tau/\tilde{\tau}_1)^{1/3\nu} \ll |s-s'| \ll (\tau_{\theta}/\tilde{\tau_1})^{1/3\nu}$ where both the exponents of $|s-s'|$ exists. 

\paragraph{\bf End-to-end distance\\}
For the case of end-to-end distance the active force dominates over the thermal noise, 
\begin{equation}
\begin{split} \langle [\R(N,t)-\R(0,t)]^2 \rangle_{\text{total}} \simeq 
2\,\, \frac{g(\nu)}{\pi}\,\, \bigl(\frac{P_b |f|\, a_h\, \overline{\tau}_1}{\zeta}\bigr)^2  \bigl(\frac{\tau_{\theta}}{\overline{\tau}_1}\bigr) \\ \bigl(1+  \frac{p\,\tau}{\tau_{\theta}} \bigr) 
\, N^{1+\nu}\end{split}
\end{equation} which is evident for large $N$. Therefore, the end-to-end distance varies as $N^{1+\nu}$ in the presence of the hydrodynamics interactions.

\subsection{Mean-squared displacement}
\subsubsection{Thermal contribution}
The contribution to the thermal part for a single segment can be obtained as
\begin{equation}
\begin{split}
\langle[\R(s,t)-\R(s,t')]^2\rangle_{\text{T}} = 6\,k_B T \int^{\infty}_{-\infty} \frac{dq}{(2\pi)}  H(q) \frac{\tilde{\tau}_1(q)}{q^2}\left(1- \right. \\ \left. e^{-\frac{|t-t'|}{\tilde{\tau}_1(q)} q^2}\right)- \Delta_{\text{T}} (q_{\text{low}})
\end{split}
\end{equation}
As $q_{\text{low}} \rightarrow 0$, $\Delta_{\text{T}} (q_{\text{low}}) \sim q^{\nu}_{\text{low}}$ which is negligible compared with the first  term on the right-hand side of the above equation.  Therefore, the thermal contribution to MSD from Eq. \ref{FTGenCorrThermal} can be obtained as 
\begin{align}
\langle [\R(s,t)-\R(s,t')]^2  \rangle_{\text{T}} &=\frac{6 k_B T}{\zeta}\tau_1 \left(a_{h}^{\frac{2}{3}}a_{\ell}^{\frac{1}{3}} \frac{\Gamma(\frac{1}{3})}{2\pi \nu}\right) \left(\frac{|t-t'|}{\tau_1}\right)^{\frac{2}{3}}
\end{align}
\subsubsection{Active force contribution}
We study the time evolution of a segment with hydrodynamic and segment-to-segment interactions. Substituting $s=s'$ in Eq. \ref{MonomerDisplaceWhole}, we obtain the below equation  
\begin{equation}
\begin{split}
\langle [\R(s,t)-\R(s,t')]^2 \rangle^{(1)}_{\text{A}}
 = 2 \tau_{\theta} P_b^2\,|f|^2 \int^{\infty}_{-\infty}\frac{dq}{[2\pi]} \frac{H^2(q)}{1-(\frac{\tau_{\theta}}{\tilde{\tau}_1(q)}q^2)^2}  \\ \bigl[\frac{\tilde{\tau}_1(q)}{q^2} \left(1-e^{-\frac{|t-t'| q^2}{\tilde{\tau}_1(q)} }\right)-\tau_{\theta}\left(1-e^{-\frac{|t-t'|}{\tau_{\theta}} }\right)\bigr] -\Delta_1(q_{\text{low}}) 
\label{RouseActiveMSD1st}
\end{split}
\end{equation}
The second term $\Delta_1 \sim q^{2\,\nu-1}_{\text{low}}$, as it is argued earlier, for finite $\tau$ and $\tau_{\theta}$, and for large $N\rightarrow \infty$, the contribution of $\Delta_1$ would be negligible compared with the first term, therefore, it is fair to drop the second term. Further, we carry out the integration and obtain
\begin{widetext}
\begin{align}
\langle[ \mathbf{R}(s,t)-\mathbf{R}(s,t')]^2 \rangle^{(1)}_{\text{A}}
\approx & \, 2\, \tau_{\theta} \frac{P^2_b |f|^2}{\zeta^2}\, \Gamma\left(\frac{2}{3}-\frac{1}{3\nu}\right)\,  \left(\frac{\overline{\tau}_1 a^2_h }{3\pi\nu} \right) \left(\frac{ \tau_{\theta}}{ \overline{\tau}_1}\right)^{\frac{1}{3}+\frac{1}{3\nu}}\,\times \begin{cases}
          \frac{3\nu}{1+\nu} \left(\frac{|t-t'|}{\tau_{\theta}}\right)^{\frac{1}{3}+\frac{1}{3\nu}}
          &\text{when $\frac{|t-t'|}{\tau_{\theta}} \gg 1$ } \\     
          \frac{1}{2}\Gamma \left(\frac{1}{3}+\frac{1}{3\nu}\right)\left(\frac{|t-t'|}{\tau_{\theta}}\right)^{2}
          &\text{when $\frac{|t-t'|}{\tau_{\theta}} \ll 1$ } \\
\end{cases}    
\end{align}
\end{widetext}
The other part of the autocorrelation of the active force can be obtained as
\begin{widetext}
\begin{align}
\langle [\R(s,t)-\R(s,t')]^2 \rangle^{(2)}_{\text{A}}
& = 2 \tau |f|^2 \, p \, P^2_b  \int^{\infty}_{-\infty}\frac{dq}{[2\pi]}\,\frac{H^2(q)}{1-(\frac{\tau}{\tilde{\tau}_1(q)}q^2)^2} \left[\frac{\tilde{\tau}_1(q)}{q^2} \left(1-e^{-\frac{|t-t'| q^2}{\tilde{\tau}_1(q)} }\right)-\tau \left(1-e^{-\frac{|t-t'|}{\tau }}\right)\right] \nonumber \\
\end{align}
\end{widetext}
Similarly, we have obtained as
\begin{widetext}
\begin{align}
\langle [\R(s,t)-\R(s,t')]^2 \rangle^{(2)}_{\text{A}}
&\approx
2\tau \frac{|f|^2}{\zeta^2} p\, P_b^2 \,\,\overline{\tau}_1 a^2_{h} \left(\frac{\tau} {\overline{\tau}_1}\right)^{\frac{1}{3}+\frac{1}{3\nu}}  \frac{\Gamma(\frac{2}{3}-\frac{1}{3\nu})}{3\pi\nu} \times
\begin{cases}
          \frac{1}{2}\Gamma \left(\frac{1}{3}+\frac{1}{3\nu}\right)\left(\frac{|t-t'|}{\tau}\right)^{2}
          &\text{when $\frac{|t-t'|}{\tau} \ll 1$ } \\
          \frac{3\nu}{\nu+1} \left(\frac{|t-t'|}{\tau}\right)^{\frac{1}{3}+\frac{1}{3\nu}}
          &\text{when $\frac{|t-t'|}{\tau} \gg 1$}          
\end{cases}
\label{GAZ_MSD_Results}
\end{align}
\end{widetext}

Due to the active force, there are three regimes as $|t-t'|$ varies from small to long time. First regime, small time $|t-t'|\ll \tau$, and second regime when $\tau_{\theta}\ll |t-t'|\ll \tau_{\text{Zimm}}$. 

\paragraph{First regime, $|t-t'|\ll \tau$:\\}
We add the terms from active force correlations ($
\langle [\R(s,t)-\R(s,t')]^2 \rangle^{(1)}_{\text{A}} + 
\langle [\R(s,t)-\R(s,t')]^2 \rangle^{(2)}_{\text{A}}$) for the regime and obtained as
\begin{equation}
\begin{split}
\langle [\R(s,t)-\R(s,t')]^2 \rangle_{\text{A}}
\simeq  P^2_b \, |f|^2 \frac{  a^2_h \, }{ \zeta^2} \,\,\overline{\tau}^{\frac{2}{3}-\frac{1}{3\nu}}_1
\frac{\Gamma(\frac{2}{3}-\frac{1}{3\nu}) \Gamma(\frac{1}{3}+\frac{1}{3\nu})}{3\pi \nu}  \\ \bigl( \tau_{\theta}^{\frac{1}{3\nu}-\frac{2}{3}} +p\, \tau^{\frac{1}{3\nu}-\frac{2}{3}} \bigr) |t-t'|^2
\end{split}
\end{equation}

\paragraph{Second regime $|t-t'| \gg \tau_{\theta}$:}
\begin{equation}
\begin{split}
\langle [\R(s,t)-\R(s,t')]^2 \rangle_{\text{A}}
\simeq 2 P^2_b \, |f|^2 \frac{  a^2_h \, }{ \zeta^2} \,\,\overline{\tau}^{\frac{2}{3}-\frac{1}{3\nu}}_1\, \frac{\Gamma(\frac{2}{3}-\frac{1}{3\nu})}{\pi (1+\nu)} \\ \bigl(\tau_{\theta}+p\, \tau \bigr) \,\, |t-t'|^{\frac{1}{3}+\frac{1}{3\nu}}
\end{split}
\end{equation}
There would be an crossover regime. The obtained scaling form is general for any feasible size exponent $\nu$. In the short time limit, the active part of the correlation varies $|t-t'|^2$. 


\section{Calculation of Correlation for Functionals}
We follow H\"{a}nggi's prescription  for the Fokker-Planck equation with colored noise \cite{Hanggi_1995}. Let us consider a functional $G[\eta(t)+x(t)]$ where $x(t)$ is a test function of $\eta(t)$. The Taylor series expansion of $G[\eta(t)+x(t)]$ in $\eta(t)$ can be expressed as
\begin{equation}
\begin{split}
G[\eta(t)+x(t)]  = G[\eta(t)] + \int dt_1 \frac{\delta G[\eta(t)]}{\delta \eta(t_1)} x(t_1)  +\frac{1}{2} \int dt_1 \\ \int dt_2 \frac{\delta^2 G[\eta(t)]}{\delta \eta(t_1) \delta \eta(t_2)} x(t_1) x(t_2)  + \dots +\frac{1}{n!} \int dt_1 \dots dt_n \\ \frac{\delta^n G[\eta(t)]}{\delta \eta(t_1) \dots \delta \eta(t_n)} \times x(t_1) \dots x(t_n) 
\end{split}
\end{equation}
When the noises are assumed to be Gaussian, the statistical correlation between any two arbitrary functional can be expressed as 
\begin{equation}
\begin{split}
\langle F[\eta] G[\eta] \rangle = \langle F[\eta]\rangle \langle G[\eta] \rangle + \sum^{\infty}_{n=1} \frac{1}{n!} \int^{t}_{t_0} \dots \int^{t}_{t_0} \\ \bigl \langle \frac{\delta^n F[\eta]}{\delta \eta(t_1) \dots \delta \eta(t_n)} \bigr \rangle  
\times \left \langle \frac{\delta^n G[\eta]}{\delta \eta(s_1) \dots \delta \eta(s_n)} \right \rangle  \\ \prod^{n}_{i=1} C(t_i-s_i) dt_i \, ds_i
\label{GenFuncCal}
\end{split}
\end{equation}
where $C(t_i-s_i)=\langle x(t_i) x(s_i)\rangle$ is the second order cumulant. 

\subsection{Calculation for white Noise}
Considering, the $F[\eta]=x(t)$ in Eq.\, \ref{GenFuncCal}, we obtain \begin{equation}
\begin{split}
  \langle x(t) G[\eta(t)+x(t)] \rangle \nonumber  = \sum^{\infty}_{n=0} \frac{1}{n!} \int dt_1 \dots \nonumber  \int dt_n  \bigl\langle \frac{\delta^n G[\eta(t)]}{\delta \eta(t_1) \dots \delta \eta(t_n)} \\ x(t_1)  \times \dots x(t_n) x(t) \bigr\rangle \nonumber  \\
\end{split}
\end{equation}
which can be expressed as 
\begin{equation}
\begin{split}
  \langle x(t) G[\eta(t)+x(t)] \rangle   = \sum^{\infty}_{n=0} \frac{1}{n!} \int dt_1 \dots  \int dt_n  \\ \bigl\langle \frac{\delta^n G[\eta(t)]}{\delta \eta(t_1) \dots \delta \eta(t_n)} \bigr\rangle \times C_{n+1}(t,t_1, \dots t_n) 
\end{split}
\end{equation}
It is better to deal the variable as scalar, therefore we consider one component of force $\mathbf{f}_T(t)$ ($\mathbf{f}_A(t)$) i,e., $f_T(t)$ ($f_A(t)$) and carry out the calculation. Thereafter we generalize for three dimensions without changing the generality. Let us first start with the thermal noise correlation $\langle f_T(t) \delta(r-R_x(t)) \rangle$, and active noise correlation $\langle f_A(t)\delta(x-R_x(t)) \rangle$, in Eq.\, \ref{ProbFunctional}. 
We obtain  
\begin{equation}
  \begin{split}
\langle \hat{f}_T(t) \delta  (x-R_x(t)) \rangle  = \sum^{\infty}_{n=1} \frac{1}{n!} \int^{t}_{t_0} dt_{1} \dots  \int^{t}_{t_0} dt_{n} \bigl\langle \hat{f}_T(t_1) \dots \hat{f}_T(t_n) \\ \, \hat{f}_T(t) \bigr\rangle  \times \bigl\langle \frac{\delta^n \delta(x-R_x(t)) }{\delta \hat{f}_T(t_1) \dots \delta \hat{f}_T(t_n)} \bigr\rangle
\label{AppenThermal}
\end{split}
\end{equation}
The statistical properties $\langle \hat{f}_T(t) \rangle=0$, and $\langle \hat{f}_T(t) \hat{f}_T(t') \rangle= 2\, k_B T N\,f_r\,\, \delta(t-t')$ for one component.
In the above equation, only the term $n=1$  contributes and terms $n \geq 2$, vanishes. Thus we now get 
\begin{equation}
\begin{split}
\langle \hat{f}_T(t) \delta(x-R_x(t)) \rangle =  \int^{t}_{t_0} dt_{1} \,\,  \langle \hat{f}_T(t_1) \hat{f}_T(t) \rangle \, \bigl\langle  \, \frac{\delta\,\delta(x-R_x(t))}{\delta \hat{f}_T(t_1)} \bigr\rangle
  \end{split}
\end{equation}
Here, we can write, 
\begin{equation}
\frac{\delta }{\delta \hat{f}_T(t_1)} \delta(x-R_x(t))  = - \frac{d}{dx} \delta(x-R_x(t)) \,\, \frac{\delta \, R_x (t)}{\delta \hat{f}_T(t_1)}
\label{ThermalNoiseFuncDiff}
\end{equation}
Using the Langevin equation described for CM, as the thermal noise $\hat{f}_T(0,t_1)$ is linearly proportional to $\R$, we obtain  $\frac{\delta \, R_x (t)}{\delta \hat{f}_T(t_1)} = \frac{\theta(t-t_1)}{N f_r}.$ Using the above equation, we obtain
\begin{equation}
\begin{split}
\langle \hat{f}_T(t) \delta(x-R_x(t))  \rangle = -2 k_B T  \int^{t}_{t_0} dt \, \delta(t-t_1) \theta(t-t_1)  \\ \frac{d}{dx} \langle \delta(x-R_x(t)) \rangle 
\end{split}
\end{equation}
where $\theta (t-t_1)=\frac{1}{2}$ if $t=t_1.$ \cite{Fox1986} We have now obtained 
\begin{equation}
\langle \hat{f}_T(t) \delta(x-R_x(t)) \rangle = - k_B T \, \frac{d}{dx} \delta(x-R_x(t))
\end{equation}

\subsection{Calculation for Active Force}
Following the above shown prescription for thermal noise, we can get 
\begin{equation}
\begin{split}
\langle  \hat{f}_A( t)\delta(r-R_x(t)) \rangle = \frac{1}{N\,f_r} \int^{t}_{t_0} dt_1 \delta(t-t_1) \int^{t_1}_{t_0} dt_2  F_A(t-t_2) \\  \times\bigl\langle \frac{\delta }{ \delta \hat{f}_A(t_2)} \delta(x-R_x(t)) \bigr\rangle 
\end{split}
\label{ColNoiseCorr1}
\end{equation}
where functional differentiation of the delta function can be obtained as
\begin{equation}
\resizebox{0.46\textwidth}{!}{$\left\langle \frac{\delta }{\delta \hat{f}_A (t_2)} \delta(x-R_x(t)) \right\rangle = - \frac{1}{N\,f_r} \theta(t-t_2)\frac{d}{d x} \langle \delta(x-R_x(t)) \rangle $}
\label{ColNoiseDeltaDiff1}
\end{equation}
Substituting Eq. \ref{ColNoiseDeltaDiff1}, in Eq. \ref{ColNoiseCorr1}, we obtain 
\begin{equation}
\begin{split}
\resizebox{0.49\textwidth}{!}{$\left \langle \hat{f}_A(t) \delta(x-R_x(t)) \right\rangle = \frac{1}{N\,f_r} \int^{t}_{t_0} dt_2 \, F_A(t-t_2) \frac{d}{d x} \langle \,\, \delta(x-R_x(t)) \rangle $}
\end{split}
\end{equation}
where $t_0$ is set to be zero, and in steady state time $t\rightarrow \infty$ is considered. Let us denote functional $F[f_T]$ either for thermal noise $\hat{f}_T(t)$ or $\hat{f}_A(t)$, and $G[f_T]=\delta(x-R_x(t))$ as the Langevin equation contains both the noise terms. Using Eq. \ref{GenFuncCal}, we obtain the correlation for one component of the thermal noise 
\begin{equation}
\langle \widehat{f}_T(t) \delta (x-R_x(t))  \rangle = - k_B T \frac{d}{dx} \langle \delta(x-R_x) \rangle
\end{equation}
where $\widehat{f}_T(0,t)$ is the Gaussian white noise. For the colored noise, we obtain 
\begin{equation} 
\langle \widehat{f}_A(t) \delta(x-R_x(t)) \rangle = - \frac{1}{3N\,f_r} \int^{t}_{t_0} dt' \,\, F_A(t-t')
\frac{d}{d x} \langle \delta(x-R_x(t)) \rangle
\end{equation}
where $F_A (|t-t'|)$ is the autocorrelation of the active force.

\end{document}